\documentclass[reprint,
superscriptaddress,
amsmath,amssymb,
apl,
]{revtex4-1}

\usepackage{graphicx}
\usepackage{dcolumn}
\usepackage{bm}
\usepackage{comment}
\usepackage{upgreek}
\usepackage[usenames,dvipsnames]{color}
\usepackage[mathlines]{lineno}
\usepackage{multirow}
\usepackage{siunitx}
\usepackage{lipsum}

\DeclareGraphicsExtensions{.pdf,.png,.jpg}

\usepackage[colorlinks]{hyperref}

\definecolor{bluegray}{RGB}{40,180,160}
\definecolor{navygray}{RGB}{110,140,170}
\definecolor{meadowgreen}{RGB}{0,128,0}
\definecolor{magenta}{RGB}{255,0,255}
\definecolor{darkred}{RGB}{139,0,0}
\definecolor{darkgreen}{RGB}{0,127,0}

\hypersetup{
urlcolor={bluegray}, 
citecolor={bluegray}, 
linkcolor={bluegray}, 
}
\usepackage{amssymb,amsmath,amsthm,enumitem}
\begin{document}

\title{Quantum non-demolition dispersive readout of a superconducting artificial atom \\ using large photon numbers}

\author{Daria~Gusenkova} 
\email{daria.gusenkova@kit.edu}

\author{Martin~Spiecker}
\thanks{First two authors contributed equally.}
\affiliation{PHI, Karlsruhe Institute of Technology, 76131 Karlsruhe, Germany}

\author{Richard~Gebauer}
\affiliation{IPE,~Karlsruhe~Institute~of~Technology,~76344~Eggenstein-Leopoldshafen,~Germany}

\author{Madita~Willsch}
\affiliation{IAS,~J\"ulich~Supercomputing~Center,~Forschungszentrum~J\"ulich,~52425~J\"ulich, Germany}

\author{Francesco~Valenti}
\affiliation{IPE,~Karlsruhe~Institute~of~Technology,~76344~Eggenstein-Leopoldshafen,~Germany}
\affiliation{PHI, Karlsruhe Institute of Technology, 76131 Karlsruhe, Germany}

\author{Nick~Karcher}
\affiliation{IPE,~Karlsruhe~Institute~of~Technology,~76344~Eggenstein-Leopoldshafen,~Germany}

\author{Lukas~Gr{\"u}nhaupt}
\affiliation{PHI, Karlsruhe Institute of Technology, 76131 Karlsruhe, Germany}

\author{Ivan~Takmakov}
\affiliation{PHI, Karlsruhe Institute of Technology, 76131 Karlsruhe, Germany}
\affiliation{IQMT,~Karlsruhe~Institute~of~Technology,~76344~Eggenstein-Leopoldshafen,~Germany}
\affiliation{National University of Science and Technology MISIS, Moscow 119049, Russia}

\author{Patrick~Winkel}
\affiliation{PHI, Karlsruhe Institute of Technology, 76131 Karlsruhe, Germany}

\author{Dennis~Rieger}
\affiliation{PHI, Karlsruhe Institute of Technology, 76131 Karlsruhe, Germany}

\author{Alexey~V.~Ustinov}
\affiliation{PHI, Karlsruhe Institute of Technology, 76131 Karlsruhe, Germany}
\affiliation{National University of Science and Technology MISIS, Moscow 119049, Russia}
\affiliation{Russian Quantum Center, Skolkovo, Moscow 143025, Russia}

\author{Nicolas~Roch}
\affiliation{Institut~Néel,~CNRS~and~Université~Joseph~Fourier,~Grenoble,~France}

\author{Wolfgang~Wernsdorfer}
\affiliation{PHI, Karlsruhe Institute of Technology, 76131 Karlsruhe, Germany}
\affiliation{IQMT,~Karlsruhe~Institute~of~Technology,~76344~Eggenstein-Leopoldshafen,~Germany}
\affiliation{Université Grenoble Alpes, CNRS, Grenoble INP, Institut Néel, 38000 Grenoble, France}

\author{Kristel~Michielsen}
\affiliation{IAS,~J\"ulich~Supercomputing~Center,~Forschungszentrum~J\"ulich,~52425~J\"ulich, Germany}

\author{Oliver~Sander}
\affiliation{IPE,~Karlsruhe~Institute~of~Technology,~76344~Eggenstein-Leopoldshafen,~Germany}

\author{Ioan~M.~Pop}
\email{ioan.pop@kit.edu}
\affiliation{PHI, Karlsruhe Institute of Technology, 76131 Karlsruhe, Germany}
\affiliation{IQMT,~Karlsruhe~Institute~of~Technology,~76344~Eggenstein-Leopoldshafen,~Germany}

\date{\today} 

\begin{abstract}
	Reading out the state of superconducting artificial atoms typically relies on dispersive coupling to a readout resonator. For a given system noise temperature, increasing the circulating photon number $\bar{n}$ in the resonator enables a shorter measurement time and is therefore expected to reduce readout errors caused by spontaneous atom transitions. However, increasing $\bar{n}$ is generally observed to also increase these transition rates. Here we present a fluxonium artificial atom in which we measure an overall flat dependence of the transition rates between its first two states as a function of $\bar{n}$, up to $\bar{n}\approx200$. Despite the fact that we observe the expected decrease of the dispersive shift with increasing readout power, the signal-to-noise ratio continuously improves with increasing~$\bar{n}$. Even without the use of a parametric amplifier, at $\bar{n}=74$, we measure fidelities of 99\% and 93\% for feedback-assisted ground and excited state preparation, respectively. 
\end{abstract}

\maketitle
Superconducting artificial atoms are among the leading platforms for the implementation of quantum information processors, due to their amenable energy spectrum and strong coupling to electromagnetic drives~\cite{devoret2007circuit, wallraff2004strong, schuster2007resolving, Baust2016Ultrastrong, yoshihara2017beyond}. These attributes enable fast single~\cite{nakamura1999coherent, Chow2010Optimized, Chen2016DRAG, zhang2020universal} and two qubit gates~\cite{niskanen2007quantum, sillanpaa2007coherent,majer2007coupling, Bialczak2011Fast}, as well as high-fidelity quantum non-demolition (QND) readout \cite{lupacscu2007quantum, Vijay2011Observation, hatridge2013quantum,  murch2013observing, Campagne2013Persistent, Vool2016Continuous}. In the circuit quantum electrodynamics (cQED) architecture~\cite{Blais2004cQED,wallraff2004strong, Blais2020circuit}, QND readout can be achieved via dispersive coupling between a readout resonator and a superconducting artificial atom. In theory, increasing the drive of the readout resonator should improve the single-shot measurement fidelity, because the integration time can be reduced to a diminishing fraction of the artificial atom's energy relaxation time. However, increasing the readout power also introduces various non-QND processes, such as dressed dephasing \cite{Boissonneault2008dressed, Boissonneault2009dephasing}, nonlinear multimode mixing in the atom-resonator Hamiltonian \cite{Malekakhlagh2020Lifetime,Petrescu2020Lifetime} and transitions between levels which become resonant at high power~\cite{Slichter2012,Sank2016}.
While quantitative agreement between theoretical models and experimental results is still missing, in practice, a tradeoff between measurement fidelity and non-QND effects is typically reached at $\bar{n}=1-15$ circulating photons in the resonator~\cite{Johnson2012Heralded, Vool2014non, Walter2017, Minev2019Catch}. This renders Josephson parametric amplifiers with near quantum limited noise essential for single-shot QND readout \cite{Castellanos-Beltran2008, Vijay2011Observation, Bergeal2010Phase}. 

\begin{figure*}[ht!]
\def\svgwidth{\textwidth}
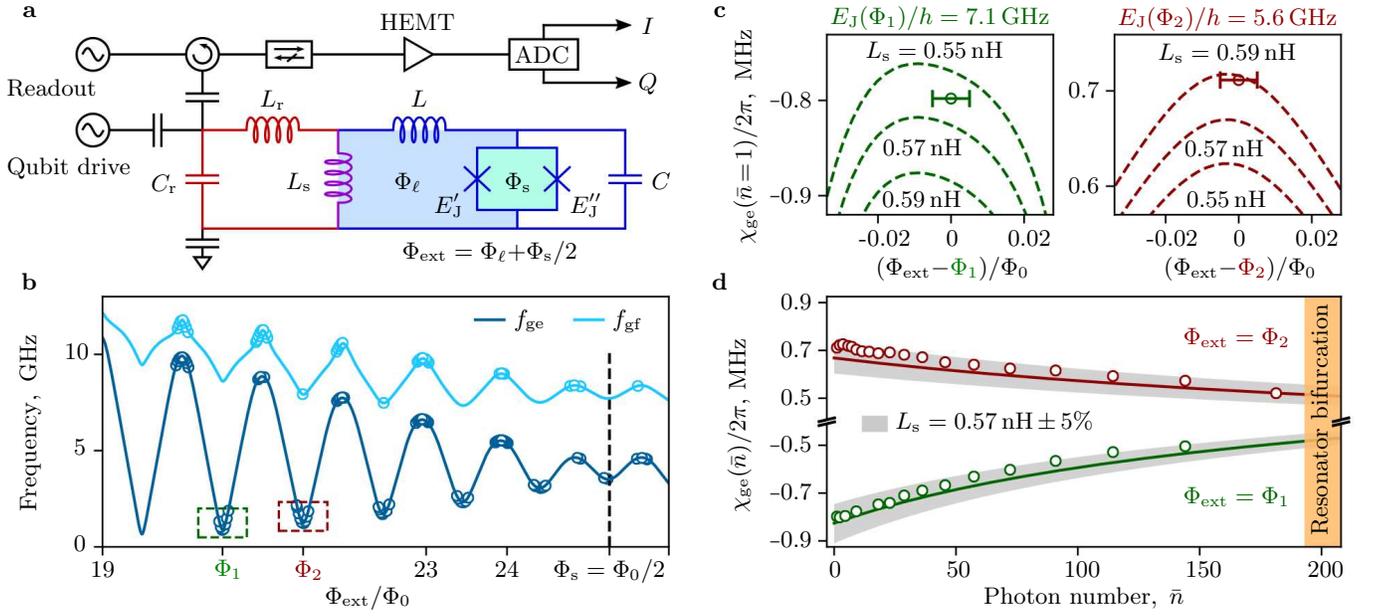 
    \caption{\textbf{Fluxonium artificial atom with in-situ tunable Josephson energy}. \textbf{(a)} Schematics of the measurement setup (black), readout resonator (red), and artificial atom (blue). The fluxonium consists of a SQUID junction, shunted by a superinductance $L = \SI{231}{\nano \henry}$, and is dispersively coupled to the resonator via a shared inductance $L_\mathrm{s}$ (violet). The capacitance $C = \SI{6.9}{\femto \farad}$ is determined by the sum of the parallel plate capacitance of the Josephson junctions and the coplanar capacitance of the electrodes. The effective Josephson energy of the SQUID, $E_\mathrm{J}$, can be tuned in-situ by the magnetic flux $\Phi_\mathrm{s}$, from its maximum $E_\mathrm{J_+}/h = \SI{24.0}{\giga \hertz}$ at $\Phi_\mathrm{s}=0$ to its minimum $E_\mathrm{J_-}/h = \SI{0.71}{\giga \hertz}$ at the SQUID frustration $\Phi_\mathrm{s} = \Phi_0/2 $ (cf. Eq.~\ref{eq_effectiveEJ}). \textbf{(b)} Measured (markers) and calculated (lines) transition frequencies between the ground state $\left| g \right\rangle$ and the first (dark blue) and second (light blue) excited states, $\left| e \right\rangle$ and $\left| f \right\rangle$ respectively, vs.~$\Phi_\mathrm{ext}=\Phi_\mathrm{\ell}+\Phi_\mathrm{s}/2$. \textbf{(c)} Dashed lines show the calculated dispersive shift $\chi_\mathrm{ge}(\bar{n}=1)$ (cf.~Eq.~\ref{eq_dispersive} and Ref.~\cite{Smith_numerics}) in the vicinity of the two frequency minima highlighted by the green and bordeaux boxes in panel~b.  The labels indicate the corresponding $L_\mathrm{s}$ values, and the markers show the measured $\chi_\mathrm{ge}(\bar{n}=1)$. Notice the sign change for $\chi_\mathrm{ge}$ between $\Phi_1$ and $\Phi_2$.  
    \textbf{(d)} Decrease of $|\chi_\mathrm{ge}|$ with increasing resonator photon number~$\bar{n}$. Markers show the measured $\chi_\mathrm{ge}$ at $\Phi_1$ (green) and $\Phi_2$ (bordeaux). Lines indicate the calculated $\chi_\mathrm{ge}$ using the values for $L$, $C$, $E_\mathrm{J}^{\prime}$, and $E_\mathrm{J}^{{\prime}{\prime}}$ extracted from the measured spectrum in Fig.~\ref{fig:1}b, and $L_\mathrm{s}=\SI{0.57}{\nano\henry}$. The grey-shaded intervals represent $L_\mathrm{s} \pm 5\%$. 
}
\label{fig:1}
\end{figure*}

In this letter we present a fluxonium artificial atom~\cite{manucharyan2009fluxonium} inductively coupled to a readout antenna, which demonstrates remarkable resilience to non-QND effects up to $\bar{n}\approx200$ readout photons. In contrast to the high power readout reported by Reed et al. in Ref.~\cite{Reed2010}, we operate the system well below its critical photon number~\cite{Blais2004cQED, Blais2020circuit} $n_\mathrm{crit}~\approx~8\cdot10^3$ in order to minimize non-QND effects. 
The main difference in our design (see Fig.~\ref{fig:1}a) compared to previous fluxonium implementations is the use of granular aluminum (grAl) for the superinductor, coupling and readout antenna inductances \cite{Grunhaupt_Fluxonium}, replacing the conventional arrays of mesoscopic Josephson junctions (JJs)~\cite{Bell2012Superinductor, Masluk2012superind}. Even though grAl can be modeled as an effective array of JJs~\cite{Maleeva2018}, it operates in a distinctly different region of the parameter space: i) its intrinsic nonlinearity can be engineered to be orders of magnitude lower compared to JJ arrays with the same inductance, and ii) its plasma frequency is about one order of magnitude higher than that of JJ superinductors, reaching values comparable to the spectral gap of the superconductor. 
These changes suppress nonlinear multimode mixing, and might explain the higher resilience of the grAl fluxonium to non-QND effects at large~$\bar{n}$ when compared with JJ array fluxoniums with similar spectra, which are susceptible to non-QND effects starting from $\bar{n}\approx2$~\cite{Vool2014non}.

The large readout dynamic range accessible in the grAl fluxonium enables the measurement of the photon number dependent dispersive shift for states $\left| g \right\rangle$ and $\left| e \right\rangle$. We show that the dispersive shift decreases with increasing readout power, as expected from the numerical diagonalization of the Hamiltonian, nevertheless, the signal-to-noise ratio (SNR) continuously improves with increasing~$\bar{n}$. At $\bar{n}=74$ we demonstrate 99\% and 93\% active state preparation fidelities for $\left|g\right\rangle$ and $\left|e\right\rangle$, respectively, without the use of a parametric amplifier. From a practical perspective, combining a strongly nonlinear spectrum and a high power QND readout provides a route for hardware efficient measurement in large quantum processors or superconducting detector arrays.

A schematic of the electrical circuit used for readout and control of the  fluxonium atom is shown in Fig.~\ref{fig:1}a. The fluxonium inductively coupled to the readout resonator is fabricated on a c-plane sapphire chip placed inside a rectangular copper waveguide \cite{Grunhaupt2018, Kou2018}. The readout resonator with a bare frequency $f_\mathrm{r0} = 1/(2\pi\sqrt{L_\mathrm{r}C_\mathrm{r}})= \SI{7.244}{\giga\hertz}$, where $L_\mathrm{r}=\SI{22.5}{\nano\henry}$, has a capacitor $C_\mathrm{r} = \SI{21.5}{\femto \farad}$ designed in the shape of a dipole antenna \cite{Grunhaupt_Fluxonium} which provides $\kappa/2\pi=\SI{1.16}{\mega\hertz}$ coupling to the microwave reflection readout setup (see Supplemental Material). All inductors shown in Fig.~\ref{fig:1}a are implemented using a $\SI{40}{\nm}$ thick grAl film deposited at room temperature, with resistivity $\rho=0.8\cdot 10^3\,\SI{}{\micro\ohm\cm}$, corresponding to a sheet inductance of $L_{\mathrm{kin}} = \SI{0.1}{\nano\henry}/\Box$. The signal reflected from the antenna is amplified by a commercial high electron mobility transistor amplifier (HEMT). 
\begin{figure}[!b]
\def\svgwidth{\columnwidth}
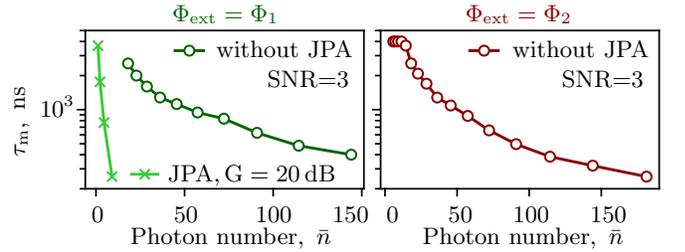
	\caption{\textbf{Measurement time $\tau_\mathrm{m}$ vs.~$\bar{n}$ for $\mathrm{SNR}=3$.} The data is acquired in a continuous wave readout, without (circles) and with (crosses) the use of a dimer Josephson junction array parametric amplifier (DJJAA)~\cite{WinkelJPA} operated at 20~dB of power gain. The $\Phi_\mathrm{ext}$ values are the same as in Fig.~\ref{fig:1}.
	}
\label{fig:2}
\end{figure}
\begin{figure*}[ht!]
\def\svgwidth{\textwidth}
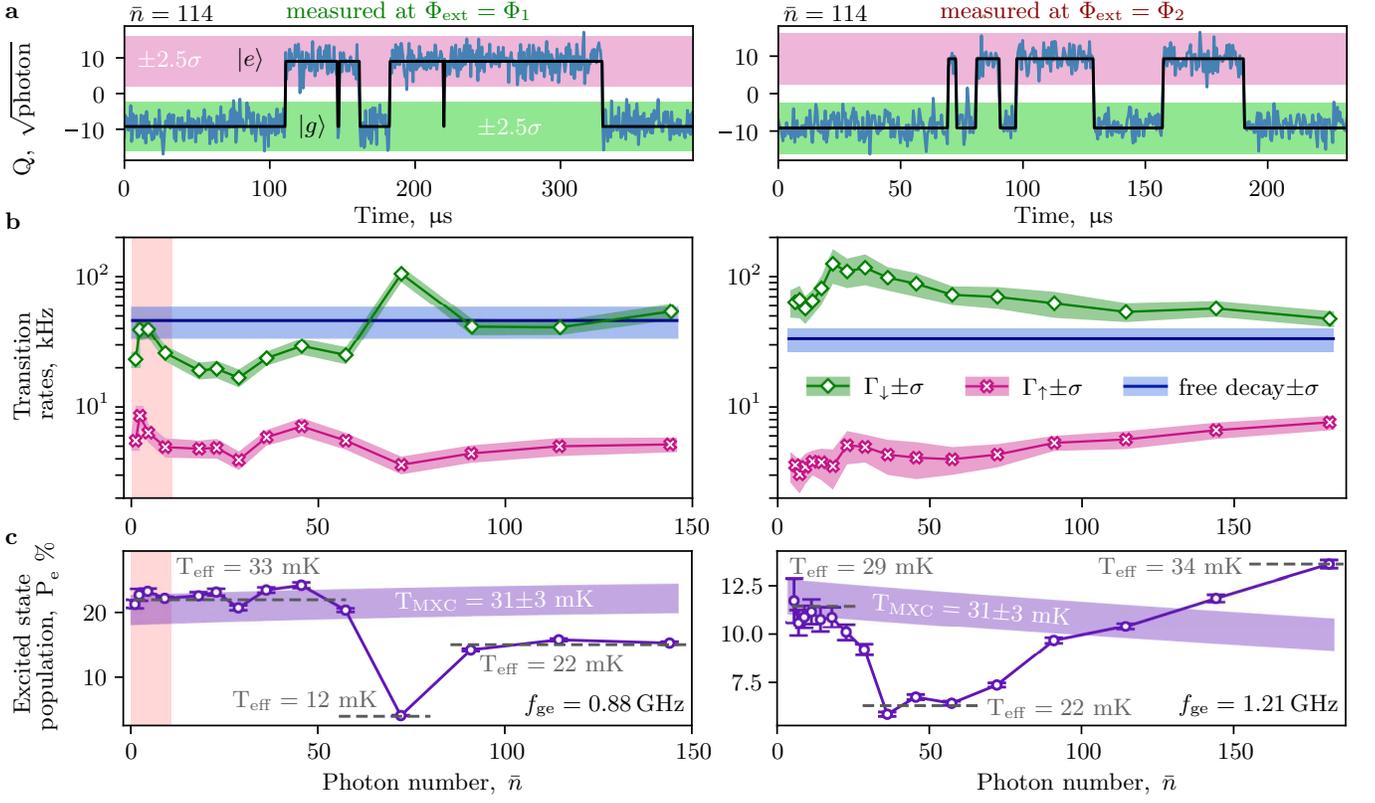
	\caption{\textbf{Measurement of quantum jumps and qubit transition rates vs. photon number.} The two columns present measurements at $\Phi_\mathrm{ext}=\Phi_1$ (left) and $\Phi_\mathrm{ext}=\Phi_2$ (right). \textbf{(a)} Typical measured time traces (blue line) of the readout resonator Q-quadrature response, without the use of a parametric amplifier, for $\bar{n}\approx114$ circulating photons. The measurement displays spontaneous quantum jumps of the qubit between the ground, $\left| g \right\rangle$, and excited state, $\left| e \right\rangle$. Shaded green and pink intervals show a $\pm 2.5 \sigma$ deviation from the centers of the corresponding distributions. The black lines indicate the qubit state estimation based on a two-point latching filter (see main text). \textbf{(b)} Fluxonium transition rates vs. $\bar{n}$ extracted from quantum jumps measurements (cf. panel a): $\Gamma_\downarrow$ (green) for $\left| e \right\rangle$ to $\left| g \right\rangle$, and $\Gamma_\uparrow$ (crimson) for $\left| g \right\rangle$ to $\left| e \right\rangle$. The blue lines indicate the measured free decay rate, defined as the inverse of the energy relaxation time. Shaded intervals correspond to $\pm$ one standard deviation. The pink shaded area in the left column presents the results obtained using the DJJAA parametric amplifier~\cite{WinkelJPA}, which allowed to detect quantum jumps at small $\bar{n}$. \textbf{(c)} Excited state population (markers) vs. $\bar{n}$, the error bars correspond to the overlap between the histograms for $\left| g \right\rangle$ and $\left| e \right\rangle$. The shaded purple interval shows the calculated thermal occupation of the $\left| e \right\rangle$ state corresponding to the mixing chamber temperature of the dilution refrigerator, $\mathrm{T_{MXC}}\!=\!31 \pm \SI{3}{\milli \kelvin}$. Note that the qubit frequency is different at the two flux biasing points, and it depends on $\bar{n}$ due to the AC-Stark effect. Grey labels corresponding to the dashed lines indicate the qubit effective temperature extracted from the measured $\left| e \right\rangle$ population, assuming thermal equilibrium.
	}
\label{fig:3}
\end{figure*}
At room temperature the output signal is digitized and decomposed into the I and Q quadratures by a custom designed field programmable gate array (FPGA) board \cite{gebauer2019state}, which allows to implement qubit state dependent feedback pulses with an instrument latency of $\SI{428}{\nano\second}$. The measurement SNR is defined as the IQ plane distance between $\left| g \right\rangle$ and $\left| e \right\rangle$ pointer states, divided by the sum of their standard deviations. 

Following the notations in Fig.~\ref{fig:1}a, the fluxonium Hamiltonian~\cite{manucharyan2009fluxonium} is $H = \frac{1}{2C} q^2 + \frac{1}{2L} \phi^2 - E_\mathrm{J}(\phi)$, 
where $\phi$ and $q$ are the node flux and charge operators, respectively, $C$ is the capacitance of the fluxonium JJ including the contribution of the leads, $L$ is the superinductance, assumed to be a linear lumped-element, and $E_\mathrm{J}(\phi)$ is the equivalent Josephson energy of the fluxonium JJ implemented using a SQUID, and therefore tunable via the external magnetic flux $\Phi_\mathrm{ext} = \Phi_\mathrm{\ell} + \Phi_\mathrm{s} / 2$: 
 \begin{equation}
    \begin{split}
    E_\mathrm{J}(\phi) = \; &\mathrm{sgn}(E_\mathrm{J_+}) \sqrt{E^2_\mathrm{J_+} + E^2_\mathrm{J_-} } \\ 
    \times & \cos{\left(\frac{2\pi}{\Phi_0}(\phi - \Phi_\mathrm{ext}) - \arctan{ \frac{E_\mathrm{J_-}}{E_\mathrm{J_+} }} \right)}.
    \end{split}
    \label{eq_effectiveEJ}
\end{equation}
Here we use the notation: $E_\mathrm{J_+}(\Phi_\mathrm{s}) =  \left(E_\mathrm{J}^{\prime} + E_\mathrm{J}^{\prime\prime} \right)\cos{\frac{\pi\Phi_\mathrm{s}}{\Phi_0}}$ and $E_\mathrm{J_-}(\Phi_\mathrm{s}) = \left(E_\mathrm{J}^{\prime} -E_\mathrm{J}^{\prime\prime} \right)\sin{\frac{\pi\Phi_\mathrm{s}}{\Phi_0}}$, where $E_\mathrm{J}^{\prime}$ and $E_\mathrm{J}^{\prime\prime}$ are the  individual Josephson energies of the two JJs in the SQUID. Fig.~\ref{fig:1}b shows the measured (markers) and fitted (lines) fluxonium spectrum for the first three levels $\left| g \right\rangle$, $\left| e \right\rangle$ and $\left| f \right\rangle$ as a function of $\Phi_\mathrm{ext}$. We obtain the optimum SNR for resolving $\left| g \right\rangle$ and $\left| e \right\rangle$ states at two minima of the fluxonium spectrum, $\Phi_\mathrm{ext}=\Phi_1$ and $\Phi_\mathrm{ext}=\Phi_2$ (see Fig.~\ref{fig:1}b), for which $|\chi_\mathrm{ge}|\approx\kappa$, where $\chi_\mathrm{ge}$ is the dispersive shift of the readout resonator frequency:
\begin{equation}
   \hbar \, \chi_\mathrm{ge}(n) = \left(E_{\left|n+1, \mathrm{e} \right\rangle} - E_{\left|n, \mathrm{e} \right\rangle} \right)
      - \left(E_{\left|n+1, \mathrm{g} \right\rangle} - E_{\left|n, \mathrm{g} \right\rangle} \right).
\label{eq_dispersive}
\end{equation}
$E_{\left|n, \mathrm{g} \right\rangle}$ and $E_{\left|n, \mathrm{e} \right\rangle}$ are the eigenenergies of the coupled fluxonium-resonator Hamiltonian. The dispersive shift is flux-dependent, and can take both negative and positive values, as shown in Fig.~\ref{fig:1}c for $\Phi_1$ and $\Phi_2$, respectively. Following a similar approach to~Ref.~\cite{Smith_numerics}, we calculate the dispersive shift by numerically diagonalizing the atom-resonator Hamiltonian, truncating the Hilbert space at 20 levels for the fluxonium and $n=220$ levels for the resonator (see Supplemental Material for the atom-resonator Hamiltonian and matrix elements of the charge and flux operators). In Fig.~\ref{fig:1}d we plot the calculated (lines) and measured (markers) $\chi_\mathrm{ge}(\bar{n})$ at $\Phi_\mathrm{ext}=\Phi_1$ and $\Phi_\mathrm{ext}=\Phi_2$. The values of $\chi_\mathrm{ge}$ were extracted from the measured phase shift of the resonator, accounting for its nonlinearity (see Supplemental Material). The numerical model predicts a decrease of $|\chi_\mathrm{ge}|$ with increasing $\bar{n}$, as observed in our measurements.

The SNR's dependence with $\bar{n}$ is the result of decreasing $\chi_\mathrm{ge}(\bar{n})$ and increasing $\sqrt{\bar{n}}$: 
\begin{equation}
    \mathrm{SNR} = \frac{\sqrt{n_\mathrm{m}(\bar{n})}}{ \sqrt{n_\mathrm{n} / 2}} \frac{2 \kappa \chi_\mathrm{ge}(\bar{n}) }{ \kappa^2 + \chi_\mathrm{ge}^2(\bar{n})},
\end{equation}
where $n_\mathrm{m} = \bar{n} \kappa \tau_\mathrm{m} / 4$ is the measurement photon number, $\tau_\mathrm{m}$ is the measurement time, and  $n_\mathrm{n}$ is the added noise photon number (see Supplemental Material). In our case, the SNR overall improves with increasing readout power, which is evidenced by the reduction of the measurement time required to obtain $\mathrm{SNR}=3$ (see Fig.~\ref{fig:2}). In combination with parametric amplification this enables the discrimination of atom states faster than the response time of the readout resonator \cite{takmakov2020minimizing}.  

We perform a continuous measurement of the qubit state with $\bar{n} = 114$ and $\mathrm{SNR}=3$, using $\tau_\mathrm{m} = \SI{480}{\nano\second}$ at $\Phi_\mathrm{ext}=\Phi_1$ and $\tau_\mathrm{m} = \SI{380}{\nano\second}$ at $\Phi_\mathrm{ext}=\Phi_2$, as shown by the blue traces in Fig.~\ref{fig:3}a. The IQ plane is rotated such that the signal is entirely contained in the Q quadrature, and we use a two-point latching filter to assign the qubit states (black line). At each time index the filter value remains latched unless the next measured point is within a $\pm2.5\sigma$ region centered on the other state. The histograms of time intervals spent in each state are fitted with an exponential function (see Supplemental Material) to obtain the transition rates $\Gamma_\uparrow$ and $\Gamma_\downarrow$, shown in Fig.~\ref{fig:3}b as a function of $\bar{n}$. Notice that $\Gamma_\downarrow$ is comparable with the measured free decay rate (shown in blue) across the $\bar{n}$ range, even though significant spikes in $\Gamma_\downarrow$ are observed at $\bar{n} = 72$  at $\Phi_\mathrm{ext}=\Phi_1$ and $\bar{n} = 18-54$ at $\Phi_\mathrm{ext}=\Phi_2$. These flux bias and $\bar{n}$ dependent $\Gamma_\downarrow$ spikes may originate from spurious resonances between readout and high order fluxonium transitions, similar to Ref.~\cite{Sank2016}. 

\begin{figure}[ht!]
\def\svgwidth{\columnwidth}
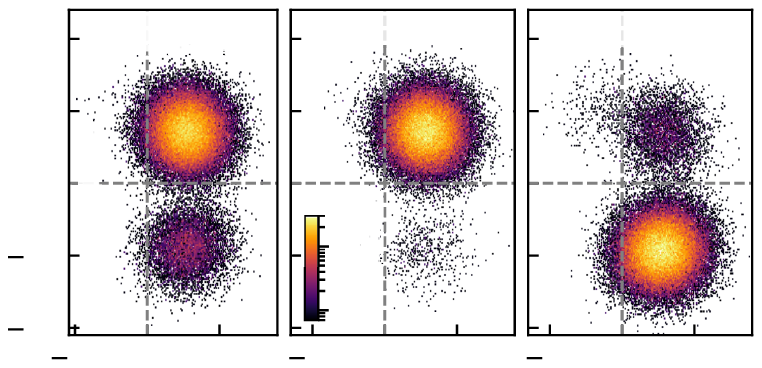
	\caption{\textbf{Fluxonium state preparation using $\bar{n} = 74$.} \textbf{(a)} Measured IQ histograms for readout duration $\tau_\mathrm{m} = \SI{560}{\nano \second}$. The Gaussian clouds correspond to steady state $\left| g\right \rangle$ and $\left| e\right \rangle$ populations measured at $\Phi_\mathrm{ext}=\Phi_2$ without using a  parametric amplifier. We extract a measurement efficiency $\eta=6\%$ corresponding to an effective noise temperature of \SI{6.0}{\kelvin} (see Supplemental Material). \textbf{(b and c)} Histograms for $\left| g\right \rangle$ and $\left| e\right \rangle$ state preparation, respectively. }
\label{fig:4}
\end{figure}

The measured transition rates are also reflected in the fluxonium population shown in Fig.~\ref{fig:3}c. At low photon number, for both flux bias points, the population of the excited state corresponds to a thermal excitation of the fluxonium levels in agreement with the cryostat temperature $\mathrm{T_{MXC}} = 31 \pm \SI{3}{\milli \kelvin}$. With increasing $\bar{n}$ the expected equilibrium population changes due to the change in the frequency of the fluxonium caused by the AC-stark shift, as indicated by the purple intervals in Fig.~\ref{fig:3}c.

Interestingly, we observe both positive and negative deviations from this trend, consistent with the measured transition rates in Fig.~\ref{fig:3}b. We can also extract the readout QND infidelity from the conditional probabilities to detect the same fluxonium state in two successive measurements: $1 - \mathcal{Q} = 1 - (P_\mathrm{g|g} + P_\mathrm{e|e}) / 2$ \cite{dassonneville2019fast,Touzard2019gated}. The QND infidelity depends on the transition rates and integration time, and it decreases with $\bar{n}$ (cf. Supplemental Material). At $\bar{n} = 114$ we measure a QND infidelity of $2-3\%$.

An essential figure of merit for our high photon number dispersive readout is the process fidelity of the active fluxonium state preparation. We use a custom-designed FPGA electronics board~\cite{gebauer2019state} to realize the measurement-based feedback (see Supplemental Material for the pulse sequences). In Fig.~\ref{fig:4}a we show the measured IQ histograms at $\bar{n} = 74$ for the fluxonium before the feedback, and in Fig.~\ref{fig:4}b and c we plot the histograms after state preparation in $\left| g\right \rangle$ and $\left| e\right \rangle$, respectively. The $\left| e\right \rangle$ state preparation histogram reveals a spurious third cloud, likely corresponding to the fluxonium second excited state $\left| f\right \rangle$. The $\approx1\%$ population of the $\left| f\right \rangle$ state extracted from IQ histograms is $\propto10^2$ times larger than expected from thermal equilibrium. The state preparation fidelities are above $90\%$ (for $\left| e\right \rangle$) and $98\%$ (for $\left| g\right \rangle$) starting from $\bar{n}\approx26$ up to $\bar{n}\approx140$ (see Supplemental Material). Using a parametric amplifier~\cite{WinkelJPA} significantly reduces the measurement time and we achieve $97\%$ fidelity for the $\left| e\right \rangle$ state preparation.

In conclusion, we demonstrated a fluxonium artificial atom, dispersively coupled to a readout resonator, for which the transition rates between the ground and excited state remain overall flat
with increasing photon number, up to $\bar{n} \approx 200$. Since the measured decrease of the dispersive shift with increasing $\bar{n}$ is slower than the SNR improvement from the stronger readout drive, the integration time required for a given SNR decreases with $\bar{n}$, reducing the QND infidelity. For $\bar{n}=74$ we achieve state preparation fidelities of 99\% for the ground, and 93\% for the excited state without the use of a parametric amplifier. We believe the stability of the transition rates as a function of $\bar{n}$ is a consequence of using grAl for the superinductor, coupling and readout
antenna inductances, instead of conventional JJ arrays, suppressing multimode mixing in the atom-resonator Hamiltonian. Future efforts will focus on the development of a quantitative model which can be employed to mitigate non-QND effects in superconducting circuits subjected to strong readout drives or parametric pumping ~\cite{Touzard2018Stabilized, lescanne2020exponential, grimm2019kerrcat}. 
 
We are grateful to A. Blais, K. Borisov, D. DiVincenzo, A. Petrescu, U. Vool and D. Willsch for insightful discussions, and to  A. Lukashenko and L. Radtke for technical support. Funding was provided by the Alexander von Humboldt foundation in the framework of a Sofja Kovalevskaja award endowed by the German Federal Ministry of Education and Research, and by the Initiative and Networking Fund of the Helmholtz Association, within the Helmholtz Future Project Scalable solid state quantum computing. RG acknowledges support by the State Graduate Sponsorship Program (LGF) and the Helmholtz International Research School for Teratronics (HIRST). AVU acknowledges partial support from the Ministry of Education and Science of Russian Federation in the framework of the Increase Competitiveness Program of the National University of Science and Technology MISIS (Grant No. K2-2020-017).

\bibliography{main}

\onecolumngrid
\vspace{5cm}
\clearpage

\section*{Supplemental Material}

In this supplemental material we provide details on the atom-resonator Hamiltonian, the measurement setup, the intrinsic and inherited nonlinearity of the readout resonator, the matrix elements of the charge and flux operators, the measurement of the photon number dependent dispersive shift $\chi_\mathrm{ge}(\bar{n})$, the signal-to-noise ratio and the  measurement efficiency $\eta$, the measured histograms of the fluxonium quantum jumps, the fluxonium total decay rate, the QND infidelity of the continuous wave measurement, the active fluxonium state preparation protocol, the fidelity and the error budget for the fluxonium state preparation.   
\vspace{1cm}

\section{Atom-resonator Hamiltonian}
Following the schematics of Fig.~1a in the main text, the Hamiltonian of the fluxonium atom coupled inductively to the readout resonator is:
\begin{equation}
    H = \frac{1}{2} \phi^T \frac{1}{L_\mathrm{r}L + L_\mathrm{r}L_\mathrm{s} + L L_\mathrm{s}} 
\begin{pmatrix}
  L + L_\mathrm{s} & -L_\mathrm{s}\\ 
  -L_\mathrm{s} & L_\mathrm{r} + L_\mathrm{s}
\end{pmatrix} \phi 
+ \frac{1}{2} q^T 
\begin{pmatrix}
  \frac{1}{C_\mathrm{r}} & \\ 
   & \frac{1}{C}
\end{pmatrix} q
- E_\mathrm{J}\left(\phi\right),
\end{equation}
where $\phi$ and $q$ are the flux and charge matrix operators, respectively. The fluxonium superinductance, readout resonator and coupling inductances are given by $L$, $L_\mathrm{r}$ and $L_\mathrm{s}$, respectively, and the capacitances of the fluxonium junction and the readout resonator are $C$ and $C_\mathrm{r}$, respectively. The  flux-dependent Josephson energy of the fluxonium junction (see Eq.~1 in the main text) is denoted as $E_\mathrm{J}(\phi)$. All inductances are considered to be linear lumped elements. \\

The eigenenergies $E_{\left|n,i\right\rangle}(\phi)$ were obtained by numerically diagonalizing the Hamiltonian matrix. We calculate the matrix elements using two different bases. The numerical diagonalization using the normal mode basis \cite{Smith_numerics} is computationally intensive and required truncation at $n=150$ levels in the resonator mode and $i = 15$ levels in the fluxonium mode. Although, this is sufficient to fit the fluxonium spectrum measured at low readout power, it does not cover the power range used in the measurement of the dispersive shift $\chi_\mathrm{ge}(\bar{n})$. The bare harmonic oscillator (Fock) basis allowed the diagonalization of the Hamiltonian matrix with larger Hilbert space dimensions ($n=220$ levels for the resonator and $i = 20$ levels for the fluxonium). This basis was used to fit $\chi_\mathrm{ge}(\bar{n})$. The difference between the dispersive shift computed in the normal and bare basis (up to $n=150$) is below $0.1\%$.

\clearpage

\section{Interferometric time-domain measurement setup}

\begin{figure*}[h!]
	\def\svgwidth{\textwidth}
	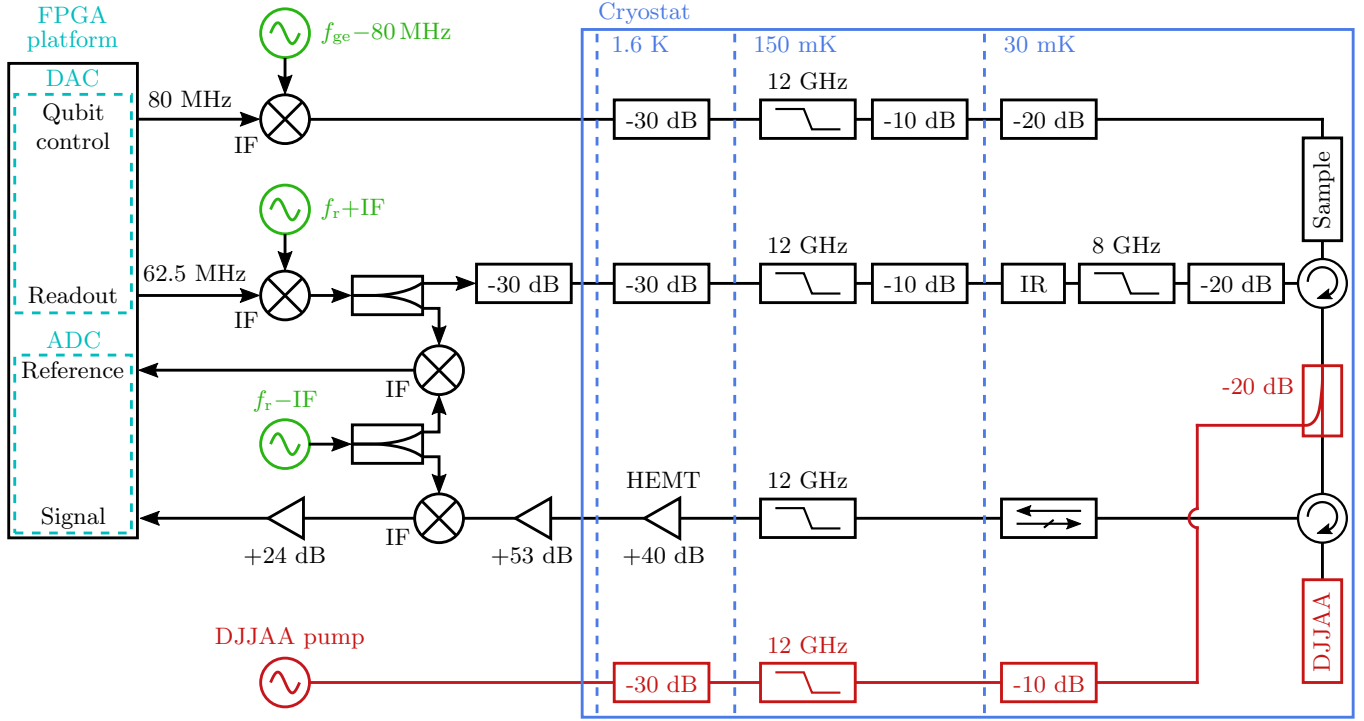
	\caption{\textbf{Interferometric time-domain measurement setup}. The digital-to-analog converter (DAC) of the  custom designed FPGA-based platform~\cite{gebauer2019state} generates waveforms for the readout and fluxonium control pulses, with IF frequencies of 62.5~MHz and 80~MHz, respectively.  The DAC waveforms are used to modulate the signals from the microwave generators (represented in green) and to convert them to the resonator frequency $f_\mathrm{r}$ and to the frequency of the fluxonium $\left| g \right\rangle$-$\left| e \right\rangle$ transition. The readout line is divided in two parts. The first part of the readout tone is directly downconverted and digitized, and it serves as reference at the analog-to-digital converter (ADC) input of the FPGA. The second part of the readout tone passes through the cryostat, where the signal interacts with the sample. The signal reflected from the sample is amplified by a cascade of cryogenic and room temperature amplifiers, downconverted, and digitized at the signal port of the ADC. The digitized \textit{signal} and \textit{reference} waveforms are used to calculate the I and Q quadratures of the readout resonator. To activate the parametric amplifier (DJJAA)~\cite{WinkelJPA}, the pump tone is applied through the microwave line shown in red color, connected to the signal line via a commercial directional coupler. When not pumped, the DJJAA can be considered as a perfect reflector.} \label{fig:9}
\end{figure*}
\clearpage

\section{Intrinsic and inherited nonlinearity of the readout resonator}\label{section_nonlinearity}

\begin{figure*}[h!]
	\def\svgwidth{\textwidth}
	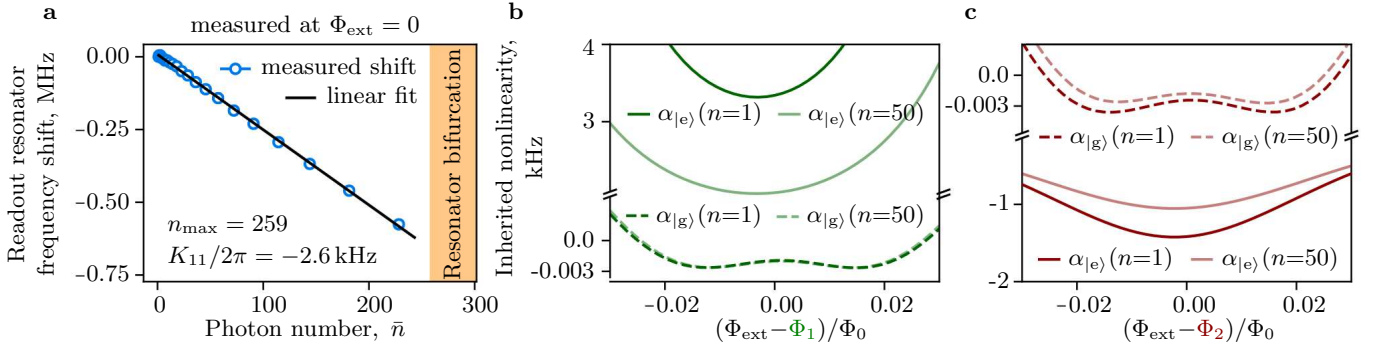
	\caption{\textbf{Intrinsic and inherited nonlinearity of the readout resonator.}  \textbf{(a)} Frequency shift of the readout resonator vs.~$\bar{n}$  (blue circle  markers) solely due to the intrinsic nonlinearity of the grAl inductor~\cite{Maleeva2018}, measured at zero external flux where the fluxonium atom is decoupled from the resonator. The black line is a linear fit. The labels indicate the self-Kerr coefficient and calculated maximal photon number at bifurcation $n_\mathrm{max} = \kappa / (\sqrt{3} K_{11})$ \cite{Eichler2014}, where $\kappa/2\pi = \SI{1.16}{\mega\hertz} $ is the resonator linewidth. \textbf{(b, c)} Nonlinearity of the resonator inherited from coupling to the fluxonium atom, calculated numerically in the vicinity of $\Phi_\mathrm{ext}=\Phi_1$ (green lines) and $\Phi_\mathrm{ext}=\Phi_2$ (bordeaux lines) from the eigenenergies of the fluxonium-resonator Hamiltonian: $h \, \alpha_{|i\rangle}(n) = \left(E_{\left|n+1, i \right\rangle} - E_{\left|n, i \right\rangle} \right) - \left(E_{\left|n, i \right\rangle} - E_{\left|n-1, i \right\rangle} \right)$, for $i = \{g, e\}$. The inherited nonlinearity is fluxonium state dependent: for the $\left| g\right \rangle$ state (dashed lines) at both flux points $\Phi_1$ and $\Phi_2$ it is in the range of few Hz, while for the $\left| e\right \rangle$ state it is orders of magnitude stronger, and it has different signs at $\Phi_1$ and $\Phi_2$. The inherited nonlinearity also depends on the photon number in the resonator: green and red lines correspond to $\alpha_{|i\rangle}(n=1)$ while faded green and faded red lines correspond to $\alpha_{|i\rangle}(n=50)$. } \label{fig:5}
\end{figure*}
\clearpage

\section{Matrix elements of the charge and flux operators}\label{matrix_elements}

\begin{figure}[h!]
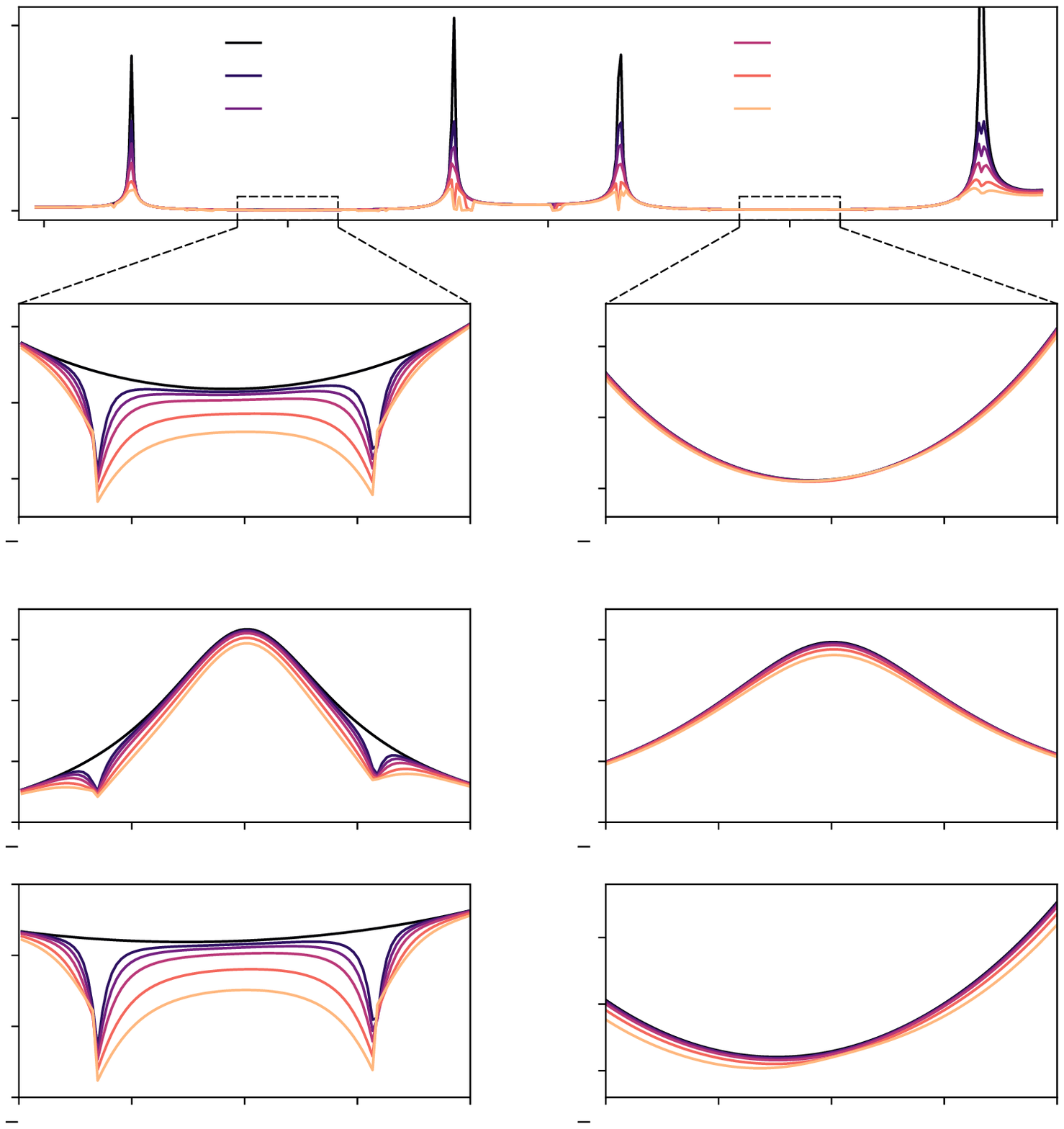
	\caption{\textbf{Absolute value of the matrix elements between states $\left| n,e \right\rangle$ and $\left| n,g \right\rangle$ vs. $\Phi_\mathrm{ext}$. } Panels \textbf{a, b, c } show the matrix elements of the readout resonator charge $q_\mathrm{r}$, atom flux $\phi_\mathrm{a}$, and atom charge $q_\mathrm{a}$ operators, respectively. The color code indicates the readout resonator photon number. The peaks of the resonator charge matrix elements (panel a) correspond to crossings between the levels $\left| 0,e \right\rangle$ and $\left| 1,g \right\rangle$, while the dips in the vicinity of $\Phi_\mathrm{ext} = \Phi_1$ correspond to crossings between the levels $\left| 1,e \right\rangle$ and $\left| 0,f \right\rangle$. }
\label{fig:15}
\end{figure}
\clearpage

\section{Measurement of the dispersive shift}

\begin{figure*}[h!]
\def\svgwidth{\textwidth}
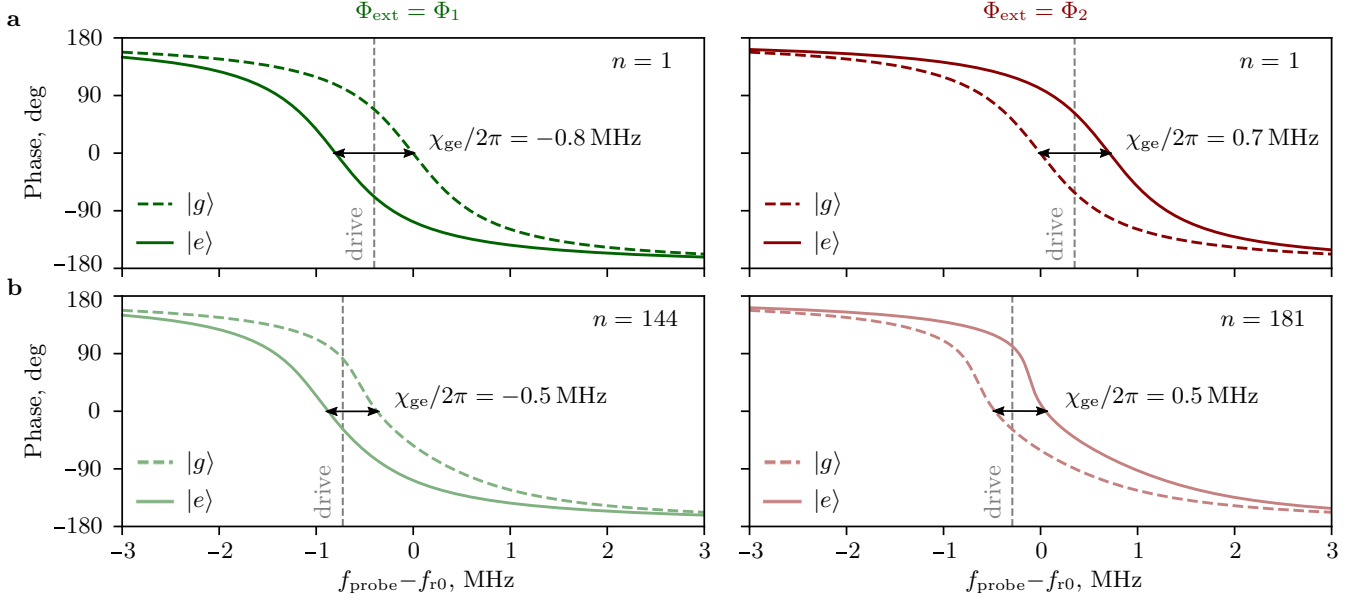
	\caption{\textbf{Phase response of the resonator for $\left| g\right \rangle$ and $\left| e\right \rangle$ fluxonium states used to extract the dispersive shift.} The phase of the signal reflected from the resonator as a function of detuning from the bare resonance frequency $f_\mathrm{r0} = \SI{7.244}{\giga\hertz}$ was calculated using input-output theory \cite{Eichler2014}. The left and right columns of the figure correspond to the external flux bias points  $\Phi_\mathrm{ext}=\Phi_1$ and $\Phi_\mathrm{ext}=\Phi_2$, respectively. The photon number was calibrated with the AC-Stark shift of the fluxonium frequency for $\bar{n}<15$, and extrapolated linearly for higher readout power values. \textbf{(a)} Phase response in the single-photon regime. The dashed and solid lines correspond to the fluxonium being in the  $\left| g\right \rangle$ and $\left| e\right \rangle$ states, respectively. The dashed grey lines show the readout drive frequency chosen to give the optimal phase separation between the IQ plane pointer states. The dispersive shift was obtained by matching the maximal phase separation between the curves for $\left| g\right \rangle$ and $\left| e\right \rangle$ states to the measured phase between the IQ plane pointer states. \textbf{(b)} Phase response at the highest photon number used in experiments, $\bar{n}=144$ at $\Phi_\mathrm{ext}=\Phi_1$ and $\bar{n}=181$ at $\Phi_\mathrm{ext}=\Phi_2$.}
\label{fig:10}
\end{figure*}

The photon number dependent dispersive shift $\chi_\mathrm{ge}(\bar{n})$ was extracted from the measured phase separation of the output $\left| \overline{n, g} \right\rangle$ and $\left|\overline{n, e} \right\rangle$ states. The phase separation was converted to a frequency shift using numerical inversion of the calculated resonator phase response (cf. Fig.~\ref{fig:10}). We calculate the phase response as a function of the probe frequency using input-output theory \cite{Eichler2014}, taking into account both intrinsic and $\bar{n}$ dependent inherited nonlinearity of the resonator (see \ref{section_nonlinearity}).\\

\clearpage

\section{IQ distributions, Signal-to-noise ratio, and Measurement efficiency}

\begin{figure}[h!]
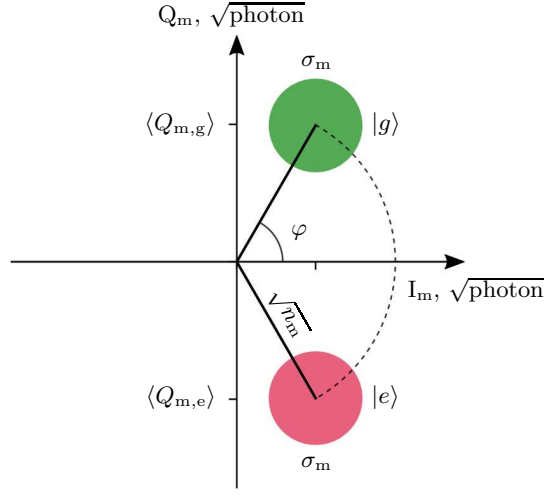
	\caption{\textbf{Schematic IQ histogram of the integrated output signal.} The I and Q quadratures are rescaled to the square root of measurement photons $\sqrt{n_\mathrm{m}}$ \cite{hatridge2013quantum, Vool2014non}. The two disks  correspond to the $\left| g\right \rangle$ and $\left| e\right \rangle$ pointer states, with identical variance $\left\langle (\Delta I_\mathrm{m})^2 \right\rangle  = \left\langle (\Delta Q_\mathrm{m})^2 \right\rangle = \sigma_\mathrm{m}^2$.}
\label{fig:13}
\end{figure}

The output field operator of a driven single-port lossless resonator obeys $\hat{a}_\mathrm{out}(w) = \frac{w - w_\mathrm{r} - i\kappa/2}{w - w_\mathrm{r} + i\kappa/2} \, \hat{a}_\mathrm{in}(w)$ \cite{Clerk2010Introduction}, here $w_\mathrm{r}$ is the resonator frequency, $\kappa$ is the  coupling rate, and  $\hat{a}_\mathrm{in}(w)$ is the input mode field amplitude operator. The dispersive coupling to the qubit results in a qubit state-dependent shift of the resonator frequency $w_\mathrm{r}  = w_\mathrm{r0} \pm \chi_\mathrm{ge}/2$. If the resonator is driven at the $w_\mathrm{r0}$ the steady state quadratures $\hat{I} = \frac{1}{2}(\hat{a}_\mathrm{out}^\dagger + \hat{a}_\mathrm{out} )$ and $\hat{Q} = \frac{i}{2}(\hat{a}_\mathrm{out}^\dagger - \hat{a}_\mathrm{out} )$ of the output signal equal 
\begin{equation}
\langle \hat{I}_\mathrm{g,e} \rangle = \frac{\kappa^2 - \chi_\mathrm{ge}^2 }{\kappa^2 + \chi_{ge}^2 } \left| \left\langle \hat{a}_\mathrm{in} \right\rangle \right| , \hspace{8pt} \langle \hat{Q}_\mathrm{g,e} \rangle = \pm \frac{2 \kappa \chi_\mathrm{ge}}{\kappa^2 + \chi_\mathrm{ge}^2 } \left| \left\langle \hat{a}_\mathrm{in} \right\rangle \right|,
\end{equation}
where the information about the qubit state is contained entirely in the Q quadrature (see Fig.~\ref{fig:13}). The output signal has to be amplified to the level accessible to the readout electronics located at room temperature. Before amplification the variance of each quadrature for the output coherent state is $\left\langle (\Delta I)^2 \right\rangle  = \left\langle (\Delta Q)^2 \right\rangle = 1/4$. In the ideal case, a quantum-limited phase-preserving amplifier adds half photon of noise \cite{Cavec1982Quantum, Clerk2010Introduction}, which results in the total  variance $\sigma_0^2 = 1/2$ for both quadratures. The quadratures of the amplified output state are obtained using the heterodyne detection \cite{Blais2020circuit} and are integrated for the measurement time $\tau_\mathrm{m}$:
\begin{equation}
Q_\mathrm{m} = \int_{t_0}^{t_0+\tau_\mathrm{m}} \left( \langle \hat{Q} \rangle + \delta Q \right) dt.
\end{equation}
Here $\langle \hat{Q} \rangle$ is the quadrature expectation value according to Eq.~2. and $\delta Q$ denotes the normally distributed random noise. The integrated value $I_\mathrm{m}$ is obtained in a similar way. Both the average value $\left\langle Q_\mathrm{m} \right\rangle = \langle \hat{Q} \rangle \tau_\mathrm{m}$ and the variance $\sigma_\mathrm{m}^2 \propto \tau_\mathrm{m}$ of the integrated signal grow linearly with the measurement time $\tau_\mathrm{m}$. We rescale the integrated  quadratures to the square root of the measurement photons $\sqrt{n_\mathrm{m}} = \sqrt{ \bar{n}\frac{\kappa}{4}\tau_\mathrm{m}}$. This makes $\sigma_\mathrm{m}$ independent of the integration time, and for a quantum-limited readout we expect  $\sigma_\mathrm{m}^2 = 1/2$. In practice, the amplification chain is not ideal, and the added noise is higher than expected in the quantum limit. The measurement efficiency is defined as $\eta = \sigma_0^2 / \sigma_\mathrm{m}^2$ \cite{hatridge2013quantum}. For the IQ histogram shown in Fig.4 of the main text we obtain $\eta = 6\%$, corresponding to $n_\mathrm{n} = 2\sigma_\mathrm{m}^2 =15.8$ noise photons per unit time and bandwidth, and an effective noise temperature $T_{\mathrm{eff}} = n_\mathrm{n} hf_\mathrm{r}/k_\mathrm{B}\approx\SI{6}{\kelvin}$. The nominal noise temperature of the HEMT amplifier (see Fig.~\ref{fig:9}) is $T \approx \SI{2}{\kelvin}$. \\

We define the signal-to-noise ratio for the qubit state measurement as
\begin{equation}
    \mathrm{SNR} = \frac{\left| \left\langle Q_\mathrm{m,g} \right\rangle -  \left\langle Q_\mathrm{m,e}  \right\rangle \right|}{\sigma_\mathrm{m,g} + \sigma_\mathrm{m,e}} = \frac{\sqrt{n_\mathrm{m}}}{ \sigma_\mathrm{m}} \sin{(\varphi)} = \frac{\sqrt{n_\mathrm{m}}}{ \sigma_\mathrm{m}} \frac{2 \kappa \chi_\mathrm{ge} }{ \kappa^2 + \chi_\mathrm{ge}^2}.
\end{equation}
Here $\varphi = \arctan \left( \left\langle Q_\mathrm{g} \right\rangle / \left\langle I_\mathrm{g} \right\rangle \right)$ is the angle formed by the pointer state associated with the qubit $\left| g\right \rangle$ state and positive real axis $I_\mathrm{m}$.\\

\clearpage

\section{Measured histograms of the fluxonium quantum jumps}

\begin{figure*}[h!]
	\def\svgwidth{\textwidth}
	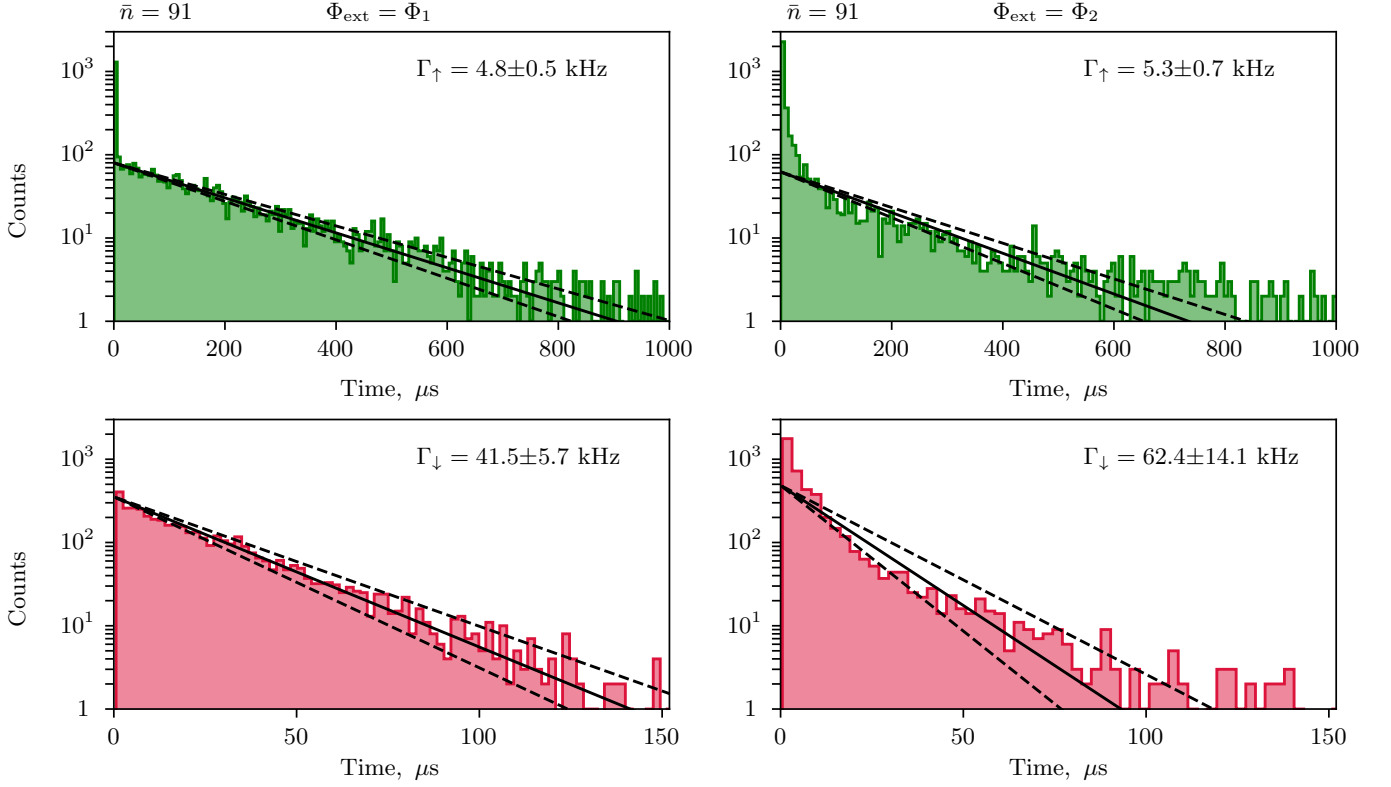
\caption{\textbf{Histograms of the fluxonium quantum jumps measured at $\bar{n} = 91$.} The left and right columns of the figure correspond to the flux biases $\Phi_\mathrm{ext}=\Phi_1$ and $\Phi_\mathrm{ext}=\Phi_2$, respectively. The time intervals spent in  the $\left| g\right \rangle$ (green) and $\left| e\right \rangle$ (pink) states were extracted from a continuous wave measurement of the fluxonium using a latching filter (see main text). The histograms contain $\sim 5\cdot10^3$ counts. In the case of uncorrelated quantum jumps the histograms should obey a Poissonian distribution $p(\tau) = \frac{1}{\bar{\tau}} e^{-\tau/\bar{\tau}}$ with the mean time $\bar{\tau}$ corresponding to the average transition rate $\Gamma = 1/\bar{\tau}$. The solid black lines show the exponential fit to the measured data, and the dashed black lines indicate the standard deviation of the extracted transition rates $\Gamma_\uparrow$ and $\Gamma_\downarrow$. Notice that the histograms of quantum jumps typically  deviate from Poissonian statistics, similarly to the results reported in Ref.~\cite{Vool2014non}. }
	\label{fig:12}
\end{figure*}

\section{Fluxonium total decay rate}

\begin{figure*}[h!]
	\def\svgwidth{\textwidth}
	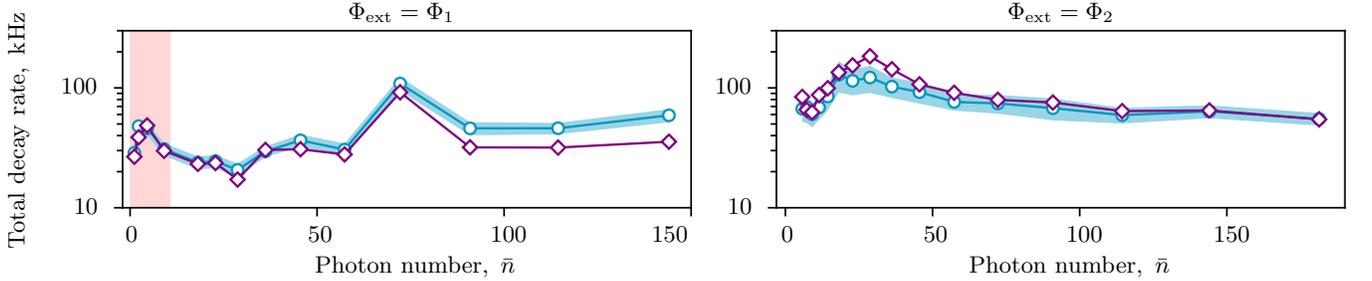
	\caption{\textbf{The fluxonium total decay rate  vs.~$\bar{n}$.} Left and right panels correspond to flux biases $\Phi_\mathrm{ext}=\Phi_1$ and $\Phi_\mathrm{ext}=\Phi_2$, respectively. The pink shaded area in the left panel indicates results obtained using the DJJAA parametric amplifier~\cite{WinkelJPA}. The circle markers show the total decay rate $\Gamma_\uparrow+\Gamma_\downarrow$ extracted from quantum jumps histograms using the two-point latching filter. The blue shaded area represents the standard deviation. The purple rhombus markers show the total decay rate calculated without using the latching filter, by averaging quantum jump traces starting from a $\pm 0.1 \sigma$ region around the center of the $\left| e \right\rangle$ state distribution. We fit the resulting $\left| e \right\rangle$ population decay using an exponential. Error bars are smaller than the marker size. Note that the two methods of calculating the  total decay rate give comparable results, validating the latching filter. } 
\label{fig:6}
\end{figure*}

\clearpage

\section{QND infidelity of a continuous wave qubit state measurement}

\begin{figure*}[h!]
	\def\svgwidth{\textwidth}
	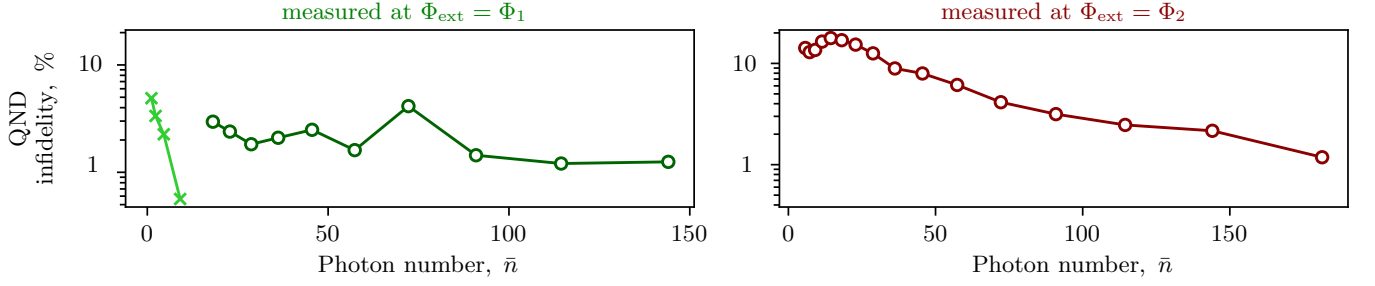
	\caption{\textbf{QND infidelity of a continuous wave qubit state measurement vs.~$\bar{n}$.} The QND infidelity is calculated from the probability to detect the same fluxonium state in two successive measurements~ \cite{dassonneville2019fast,Touzard2019gated}: $1 - \mathcal{Q} = 1 - (P_\mathrm{g|g} + P_\mathrm{e|e}) / 2$. The integration time is chosen for each photon number according to Fig.~2 in the main text. The QND infidelity generally decreases with the photon number, directly reflecting the shorter integration time required for $\mathrm{SNR}=3$.} 
	\label{fig:6}
\end{figure*}

\section{Pulse sequences used for the fluxonium state preparation}
\begin{figure}[h!]
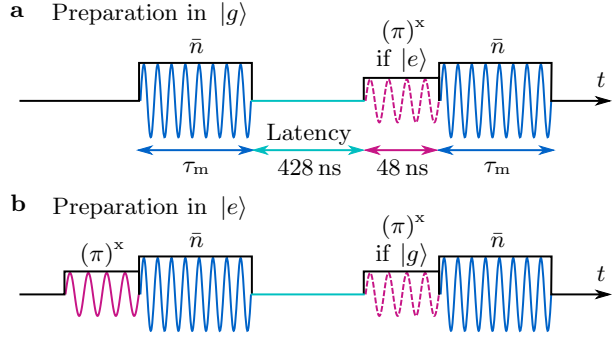
	\caption{\textbf{Pulse sequences used for the fluxonium state preparation.} \textbf{(a)} Preparation in the $\left| g\right \rangle$ state starts with a readout pulse of duration $\tau_\mathrm{m}$ (shown in blue color), which populates the readout resonator with $\bar{n}$ circulating photons. Using the measured reflected signal, the fluxonium state is evaluated on-the-fly with a custom designed FPGA-based electronics board~\cite{gebauer2019state}; the data processing takes 428~ns (the board latency time). A conditional $\pi$~pulse (shown in crimson dashed) is applied if the fluxonium was measured in the $\left| e\right \rangle$ state, otherwise the final readout pulse is played immediately after the state evaluation. The fidelity of state preparation is measured by the final readout, of the same duration and amplitude as the starting readout pulse. \textbf{(b)} Preparation in the excited state starts from the $\pi$~pulse (shown in crimson solid) which inverts the thermal population of the fluxonium $\left| g\right \rangle$ and $\left| e\right \rangle$ states. The rest of the pulse sequence is analogous to the one in panel a, with the $\pi$~pulse conditioned on the $\left| g\right \rangle$ state. }
\label{fig:8}
\end{figure}

\clearpage

\section{Fidelity of the fluxonium state preparation}
\begin{figure}[h!]
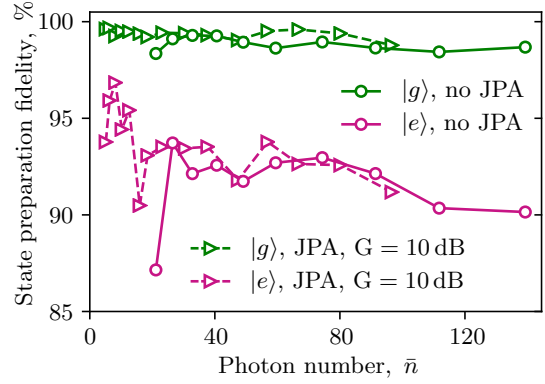
	\caption{\textbf{Fidelity of the fluxonium state preparation vs. photon number, at $\Phi_\mathrm{ext}=\Phi_2$.} The pulse sequences are presented in Fig.~\ref{fig:8}. The measurements were made without the use of a parametric amplifier (green and crimson circle markers for $\left| g\right \rangle$ and $\left| e\right \rangle$ respectively), and using a parametric amplifier (green and crimson triangular markers for $\left| g\right \rangle$ and $\left| e\right \rangle$ respectively) operated at a power gain of 10~dB. The fidelity of the state preparation is defined as the target state occupation after the feedback.}
\label{fig:11}
\end{figure}

\section{Error budget for the fluxonium state preparation}
\begin{figure*}[h!]
\def\svgwidth{\textwidth}
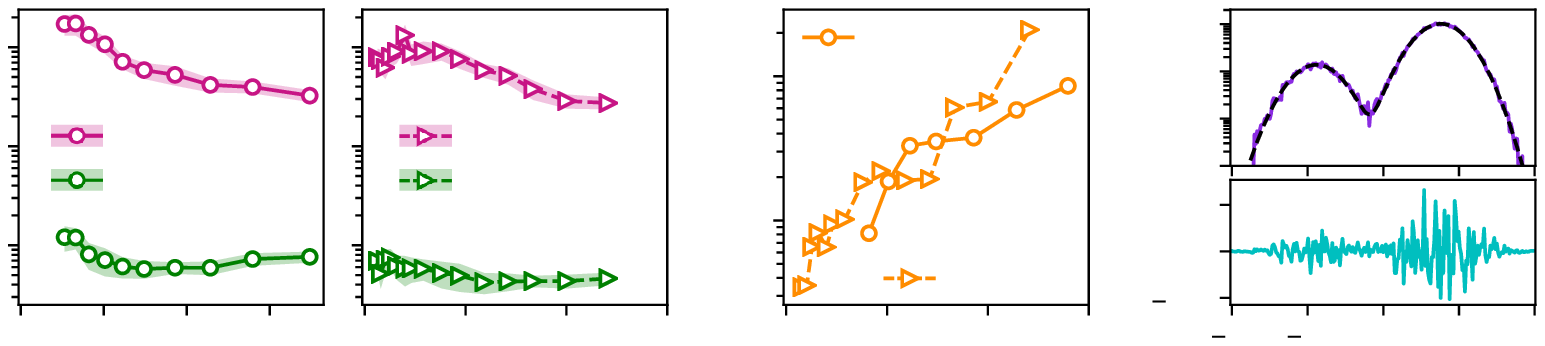
	\caption{\textbf{Error budget for the fluxonium state preparation, at $\Phi_\mathrm{ext}=\Phi_2$.} \textbf{(a)} Preparation errors due to fluxonium transitions during the measurement and the FPGA latency time, calculated from the rates $\Gamma_\uparrow$ and $\Gamma_\downarrow$ for the ground state (green markers) and the excited state (crimson markers), respectively. Left and right panels correspond to measurements without and with the use of a parametric amplifier operated at a power gain of 10~dB. The color shaded areas indicate the uncertainty corresponding to the measured standard deviation of the free decay rate for the $\left| g \right\rangle$-$\left| e \right\rangle$ transition versus time, during an interval of several hours. \textbf{(b)} $\left| f \right\rangle$ state population extracted from the measured IQ-distributions for the  $\left| e \right\rangle$ state preparation (cf. Fig~4c in main text). The lines are connecting the markers. The circle and triangular markers show results obtained without and with the use of a parametric amplifier, respectively. These values place a lower bound for the $\left| e \right\rangle$ state preparation error.
	\textbf{(c)}~Fit error for the measured Q-quadrature distribution. In the top panel, the purple line shows the distribution of counts for $5\times10^5$~single-shot readout outcomes, measured at $\bar{n} = 74$ without a parametric amplifier. The black dashed line is a double-Gaussian fit. The overlap between two Gaussian distributions imposes a lower limit for the preparation error of 1\%. The cyan line in the lower panel indicates the difference between the measured  Q-quadrature distribution and the fit. Notice that the standard deviation $\sigma$ is less than 0.1\% of the total number of counts.}
\label{fig:7}
\end{figure*}

\clearpage

\end{document}